\documentclass[
 superscriptaddress,
nofootinbib,
 amsmath,amssymb,
 twocolumn,
 aps,
 prd,
longbibliography
]{revtex4-2}

\usepackage[utf8]{inputenc}
\usepackage{geometry}
\usepackage{float}
\usepackage{graphicx,xcolor}
\usepackage{enumitem}
\usepackage{braket}
\usepackage{url} 
\usepackage[colorlinks=true]{hyperref}
\usepackage{amsfonts,amssymb}
\usepackage{amsmath, mathtools}

\usepackage[normalem]{ulem}

\usepackage{tikz}
\usetikzlibrary{arrows.meta}

\usetikzlibrary{positioning}
\usepackage[edges]{forest}
\usepackage{enumitem}
\usepackage{braket}
\usepackage{amsfonts,amssymb}
\usepackage{amsmath, mathtools}
\usepackage{caption,subcaption}
\usepackage{comment}

\usepackage{amsthm}

\begin{document}

\title{A formula for the overlap between Generalized Coherent States of any rank one simple Lie algebra}
\author{Nicola Pranzini}
\email{nicola.pranzini@helsinki.fi}
\address{Department of Physics, P.O.Box 64, FIN-00014 University of Helsinki, Finland}
\address{QTF Centre of Excellence, Department of Physics, University of Helsinki, P.O. Box 43, FI-00014 Helsinki, Finland}
\address{InstituteQ - the Finnish Quantum Institute, Finland}

\begin{abstract}
We provide a formula for computing the overlap between two Generalized Coherent States of any rank one simple Lie algebra. Then, we apply our formula to spin coherent states (i.e. $\mathfrak{su}(2)$ algebra), pseudo-spin coherent states (i.e. $\mathfrak{su}(1,1)$ algebra), and the $\mathfrak{sl}(2,\mathbb{R})$ subalgebras of Virasoro. In all these examples, we show the emergence of a semi-classical behaviour from the set of coherent states and verify that it always happens when some parameter, depending on the algebra and its representation, becomes large.
\end{abstract}

\maketitle
\section{Introduction}
Generalized Coherent States (GCS) are a set of quantum states of paramount importance in many fields of theoretical and experimental physics~\cite{ZhangFG90, Perelomov86, Combescure21}.

On the theoretical side, GCS were introduced as the generalization of Bosonic Coherent States (BCS), defined for studying the classical-like features of the quantum harmonic oscillator and the electromagnetic field~\cite{Schrodinger26, Klauder60, Glauber63}. In consequence, GCS have been used as a way to understand the semi-classical regime of various quantum systems, as well as more general features of the quantum-to-classical crossover~\cite{ZurekEtAl93, PazZ93, Zurek03, FotiEtAl19, CoppoEtAl20} and quantum chaos~\cite{DAlessioEtAl16, CaputaEtAl21, Patramanis22}. Moreover, GCS allow studying the classical limit of various QFTs~\cite{Zhang99, Yaffe82}, quantum and classical-quantum descriptions of gravity~\cite{LaytonEtAl23, LaytonO23}, and the emergence of time in quantum mechanics via the Page and Wootters mechanism~\cite{PageW83, FotiEtAl21}. More recently, GCS have been used to describe semi-classical states in CFT~\cite{Guo15, Hollands20, PalEtAl22}, and have been proven to have a well-defined gravitational dual description via the Holographic principle~\cite{CaputaG23}. Other applications to quantum gravity include the crucial case of the thermofield double state~\cite{ChapmanEtAl19, HartmanM2013}, which connects the problem of studying GCS and their classical limit to that of uncovering the semi-classical behaviour of quantum black holes~\cite{Coppo23} and the related information paradox~\cite{Israel76, Maldacena03, MaldacenaS13}.

On the experimental side, BCS and GCS are crucial tools of quantum optics and quantum information science in the lab~\cite{RosasOrtiz19}. These states appear in a wide range of applications, spanning from precision measurements~\cite{Mihalcea23} to implementing quantum communication and cryptographic schemes~\cite{WeedbrookEtAl12, AdessoEtAl14}. Moreover, assessing the fidelity between experimentally accessible coherent states (most often in quantum optics~\cite{LiangEtAl19}) enables studying information-theoretic quantities and protocols from a practical viewpoint~\cite{CerfEtAl00, HammererEtAl05, TakeiEtAl05}. In many of these applications, it is common to use squeezed states of light~\cite{Walls83}. By noticing that squeezed states are just a realization in the bosonic framework of the so-called pseudo-spin GCS, and that these are also employed in constructing the above-mentioned thermofield double states of CFTs, one can establish an often overlooked link between the tools of quantum gravity and practical experimental applications. Finally, some authors recently proposed using GCS to experimentally test low-energy quantum gravity phenomena~\cite{KumarP20, Pedram13}.

This article presents a formula for the overlap between any two coherent states of a rank one Lie algebra admitting triangular decomposition. For any given GCS construction, our result only depends on the algebra's structure constants and on the action of the algebra on the lowest weight vector used for building the GCS. Therefore, our formula has broad applicability and can easily be used in many settings, both experimental and theoretical. Moreover, since our formula distillates the essential features of the Lie algebra and GCS construction relevant for computing the overlap between two states, it paves the way for uncovering and studying the model-independent features of the semi-classical behaviour of quantum systems and their quantum-to-classical crossover.

The remainder of this work is organized as follows. In Sec.~\ref{sec:2}, we review some necessary aspects of the theory of Lie groups and algebras, which are then used in Sec.~\ref{Generalized:Coherent:States} to summarize the group-theoretic construction of GCS. Next, in Sec.~\ref{s.rank_one_overlaps}, we present and prove our formula. Finally, in Sec.~\ref{sec:5}, we apply it to  $\mathfrak{su}(2)$, $\mathfrak{su}(1,1)$, obtaining well-knonw results, and to the $\mathfrak{sl}(2,\mathbb{R})$ subalgebras of Virasoro. For all these examples, we use our formula to show the emergence of a semi-classical behaviour for the set of coherent states. In Sec.~\ref{sec:6}, we briefly recap our results and discuss possible future applications and developments.

\section{Preliminaries}
\label{sec:2}
We start by briefly reviewing some necessary tools of the theory of Lie algebras~\cite{BarutR86, Hall03}. We consider a Lie group $G$ of dimension $d$, and associate it with its finite-dimensional complex (real) Lie algebra $\mathfrak{g}$ over $K=\mathbb{C}$ ($\mathbb{R}$). The elements of the Lie group and those of the Lie algebra are related by the exponential map in the physicist's convention, i.e.
\begin{equation}
    exp:g\mapsto G=e^{ig}~.
    \label{e.exp_map}
\end{equation}
Given a basis $\{X_i\}$ of $\mathfrak{g}$, the structure constants of the algebra are the anti-symmetric coefficients $c_{ijk}$ defined by
\begin{equation}
[X_i , X_j ] =\sum_k c_{ijk} X_k~.
\label{structure_constants}
\end{equation}
A subalgebra $\mathfrak{i}\subset \mathfrak{g}$ is said to be an \textit{ideal} of $\mathfrak{g}$ if 
\begin{equation}
[g, k] \in \mathfrak{i}~,\forall g\in\mathfrak{g},~\forall k\in\mathfrak{i}~.
\end{equation}
In this article, we are interested in simple and semi-simple Lie algebras, which are defined as follows: a non-abelian Lie algebra is simple if its only ideals are $\{0\}$ and $\mathfrak{g}$ itself; a Lie algebra is semi-simple if can be expressed as a direct sum of simple subalgebras. Complex semisimple Lie algebras are relevant for our discussion because they allow the so-called triangular decomposition over the Cartan-Weyl basis~\cite{MoodyP95}. 

Given a complex semisimple Lie algebra $\mathfrak{g}$, we select the maximal subalgebra $\mathfrak{h}$ (often called Cartan subalgebra) such that its adjoint action is diagonalizable. We call rank of $\mathfrak{g}$ the dimension of $\mathfrak{h}$ and root of $\mathfrak{g}$ an element $\alpha$ of $\mathfrak{h}^*$ such that the set
\begin{equation}
\mathfrak{g}_\alpha=\{g\in\mathfrak{g} ~|~ ad_h(g)=\alpha(h)g,~\forall h\in\mathfrak{h}\}
\end{equation}
is not empty. We denote with $\Phi$ the collection of all roots. It is worth noting that $\dim(\mathfrak{g}_\alpha)=1$ for every root and that roots come in pairs, meaning that if $\alpha\in\Phi$, then $-\alpha\in\Phi$ and no other nonzero multiple of $\alpha$ is a root. Therefore, $\Phi$ admits the decomposition
\begin{equation}
\Phi=\Phi^+\cup\Phi^-~,
\end{equation}
where $\Phi^\pm$ respectively are the sets of positive and negative roots. Moreover, for every root $\alpha\in\Phi$ there exists unique element $t_\alpha$ in $\mathfrak{h}$ such that
\begin{equation}
(h,t_\alpha)=\alpha(h),~\forall h\in\mathfrak{h}~,
\end{equation}
where $(\cdot,\cdot)$ is the Killing form\footnote{The Killing form is the application $(\cdot,\cdot):\mathfrak{g}\times\mathfrak{g}\rightarrow K$ that acts on the elements of the Lie algebra as $(X,Y)=\mathrm{Tr}(ad_X ad_Y)$, where $ad_X$ is the adjoint action defined by $ad_X(Y)=[X,Y]$.} on $\mathfrak{g}$. With these definitions, it is possible to find the so-called Cartan-Weyl basis $\{h_i,g_\alpha\}$ of $\mathfrak{g}$, whose elements satisfy the commutation rules
\begin{equation}
\begin{dcases}
[h_i,h_j]=0\\
[g_\alpha,g_{-\alpha}]=(g_\alpha,g_{-\alpha})t_\alpha\\
[h_i,g_{\alpha}]=\alpha(h_i) g_\alpha,\\
[g_\alpha,g_\beta]=c_{\alpha\beta}g_{\alpha+\beta}
\end{dcases}
\label{CartanWeylBasis}
\end{equation}
where $t_\alpha\in\mathfrak{h}$ and $c_{\alpha\beta}\neq 0$ only if $\alpha+\beta\in\Phi$. This provides the triangular decomposition of the given algebra as
\begin{equation}
\mathfrak{g}=\mathfrak{h}\underset{\alpha\in\Phi^+}{\bigoplus}(\mathfrak{g}_\alpha\oplus\mathfrak{g}_{-\alpha})~.
\label{e.triangular}
\end{equation}
In particular, any complex simple rank one Lie algebra admits triangular decomposition over the three elements $\{g_-,h,g_+\}$ such that 
\begin{equation}
\begin{dcases}
[g_+,g_-]=\beta h\\
[h,g_\pm]=\alpha_\pm g_\pm,\\
\end{dcases}~.
\label{e.brackets}
\end{equation}
The formula we provide in the following holds for this class of Lie algebras and their real forms.


\section{Generalized Coherent States}
\label{Generalized:Coherent:States}
This section reviews the group-theoretic construction of the (Perelomov) Generalized Coherent States (GCS)~\cite{Perelomov77, Perelomov86, ZhangFG90}, which is a generalization of the well-known bosonic coherent states~\cite{Glauber63}.

\subsection{Definition of the GCS}
Let $R = (\Pi, \mathcal{H})$ with $\Pi:G\rightarrow GL(\mathcal{H})$ be a unitary irreducible representation over $\mathcal{H}$ of a Lie group $G$ with associated algebra $\mathfrak{g}$, and $\ket{\psi_0} \in \mathcal{H}$ a reference state vector. The choice of $\ket{\psi_0}$ selects the so-called stabilizer subgroup $H\subset G$ as the one whose elements leave the reference state vector invariant up to a phase factor. Defining the set $\{\ket{\psi_g}\}$ as the collection of elements
\begin{equation}
\ket{\psi_g} = \pi(g)\ket{\psi_0}~g\in G~,
\end{equation}
it is straightforward that all vectors $\ket{\psi_{g}}$ differing only by a phase factor are generated by the same element in the coset $G/H$. Then, a Generalized Coherent State $\ket{x}$ is defined as
\begin{equation}
\ket{x} \equiv  \Pi(x)\ket{\psi_0}~,~x\in G/H~;
\end{equation}
consequently, different elements of $G$ in the same equivalence class specified by $G/H$ give the same coherent state, times a physically irrelevant phase factor. Following the convention established for bosonic coherent states, the operators $ \Pi(x)$ are sometimes called displacement operators and denoted by the uppercase letter $D$.

As it is clear, the definition of a set of GCS depends on three ingridients~\cite{ZhangFG90}: the group $G$, the representation $R$ (and, more specifically, the Hilbert space $\mathcal{H}$), and the reference state vector $\ket{\psi_0}$. In the following, we will always consider the case in which $\ket{\psi_0}$ is an extremal state annihilated by a maximal subset of $\mathfrak{g}$, meaning that we choose it to be either the Highest (HWV) or Lowest Weight Vector (LWV) of the representation, respectively denoted by $\ket{h_0}$ and $\ket{l_0}$.

\subsection{GCS as a differential manifold}
The elements of $G/H$ are in a one-to-one correspondence with the points of a manifold $\mathcal{M}$~\cite{Lee13}. Moreover, since each element of $G/H$ corresponds to a GCS, we can extend the correspondence to identify each GCS with a point of $\mathcal{M}$. In particular, let us assume $\mathfrak{g}$ admits a triangular decomposition \eqref{e.triangular} (e.g. $\mathfrak{g}$ is semi-simple), and suppose $\mathfrak{h}$ is associated with the stabilizer subgroup $H$. Given the Cartan-Weyl basis (\ref{CartanWeylBasis}), we can choose the algebra's representation $(\pi,\mathcal{H})$ such that the operators $\pi(h_i)=\hat{H}_i$ are Hermitian and diagonal, and the so-called shift operators $\pi(g_\alpha)=\hat{E}_\alpha$ satisfy
\begin{equation}
\hat{E}_\alpha^{\dagger}=\hat{E}_{-\alpha}~.
\end{equation}
Therefore, we can write any element in $G/H$ as
\begin{equation}
D(\Omega) =e^{\sum_{\alpha}\left(\Omega_\alpha g_\alpha-\Omega^*_{\alpha}g_{- \alpha}\right)}~,
\label{e.shift}
\end{equation}
where $\Omega_\alpha \in \mathbb{C}$, and the action of the displacement operators on the reference state as
\begin{equation}
\ket{\Omega} = \hat{D}(\Omega)\ket{\psi_0}=e^{\sum_{\alpha}\left(\Omega_\alpha\hat{E}_{\alpha}-\Omega^*_{\alpha}\hat{E}_{- \alpha}\right)}\ket{\psi_0}~.
\end{equation}
The correspondence between the set of GCS and $\mathcal{M}$ is now evident: every GCS can be written in terms of a collection of $m=\dim(G/F)/2$ complex numbers, that are recognised as coordinates of the points $p_\Omega\in\mathcal{M}$.

Furthermore, $\mathcal{M}$ inherits a metric from $G/H$, which can be derived by the second order derivatives of the logarithm of 
\begin{equation}
N(\tau)=\bra{\tilde{\tau}}\tilde{\tau}\rangle~,
 \label{Fcosetspace}
\end{equation}
where the vectors $\ket{\tilde{\tau}}$ are the unnormalized coherent states~\cite{ZhangFG90}. When the GCS are built from the LWV, $\ket{\tilde{\tau}}$ is
\begin{equation}
    \ket{\tau}=N(\tau)^{-\frac{1}{2}}e^{\sum_\alpha\tau_\alpha \hat{E}_\alpha}\ket{l_0}\equiv N(\tau)^{-\frac{1}{2}}\ket{\tilde{\tau}}~,
    \label{e.unnormalized_GCS}
\end{equation}
where the GCS $ \ket{\tau}$ is obtained from $ \ket{\Omega}$ by the BCH formula
\begin{equation}
    \hat{D}(\Omega) =e^{\sum_\alpha\tau_\alpha \hat{E}_{\alpha}}e^{\sum_i\eta_i \hat{H}_i}e^{-\sum_\alpha\tau^*_\alpha \hat{E}_{-\alpha}}
\end{equation}
and the fact that the $\hat{E}_{-\alpha}$ annihilate the LWV of the chosen representation. Similarly, when the GCS are built from the HWV, $\ket{\tilde{\tau}}$ is obtained by the expression
\begin{equation}
    \ket{\tau}=N(\tau)^{-\frac{1}{2}}e^{-\sum_\alpha\tau^*_\alpha \hat{E}_{-\alpha}}\ket{h_0}\equiv N(\tau)^{-\frac{1}{2}}\ket{\tilde{\tau}}~,
\end{equation}
where we used the BCH formula
\begin{equation}
    \hat{D}(\Omega) =e^{-\sum_\alpha\tau^*_\alpha \hat{E}_{-\alpha}}e^{-\sum_i\eta_i \hat{H}_i}e^{\sum_\alpha\tau_\alpha \hat{E}_{\alpha}}
\end{equation}
and the fact that the $\hat{E}_{\alpha}$ annihilate the HWV of the chosen representation. An example of how these unnormalized states can be constructed and the related change of variables from $\Omega$ to $\tau$ can be found in App.~\ref{app:Tau-Omega}. From the metric, one can obtain a measure and a symplectic structure for the coset space, hence promoting $\mathcal{M}$ to a phase space with well-defined Poisson brackets~\cite{Perelomov86}. In turn, this gives the GCS a classical-like structure, which can sometimes be used to obtain a well-defined classical limit in both the finite~\cite{CoppoEtAl22, BalazsP23} and infinite Hilbert space dimension regimes~\cite{Yaffe82, FotiEtAl19, CoppoEtAl20}. While the manifold's Riemannian and symplectic structures play no role in our analysis, the function \eqref{Fcosetspace} and unnormalized GCS \eqref{e.unnormalized_GCS} will be crucial in the following. Finally, it is easy to show that GCS are normalized but non-orthogonal, and they resolve the identity on $\mathcal{H}$ (i.e. GCS are overcomplete).

\section{GCS overlaps for rank one Lie algebras}
\label{s.rank_one_overlaps}
Let $\mathfrak{g}$ be a rank one Lie algebra with Weyl-Cartan basis $\{g_-,h,g_+\}$ and commutation relations \eqref{e.brackets}. Once we select a representation $r=(\pi,\mathcal{H})$ such that 
\begin{equation}
    \pi(g_\pm)=\hat{K}_\pm~~,~~\pi(h)=\hat{K}_0~,
    \label{e.Rep_LWV}
\end{equation}
$\hat{K}_+^\dagger=\hat{K}_-$, and perform the GCS construction from a LWV satisfying
\begin{equation}
    \hat{K}_0\ket{l_0}=\nu_0\ket{l_0}~,~\hat{K}_-\ket{l_0}=0~,
    \label{e.HWV}
\end{equation}
we can build all coherent states as 
\begin{equation}
\ket{\tau}=N^{-1/2}(\tau)e^{\tau\hat{K}_+}\ket{l_0}\equiv N^{-1/2}(\tau)\ket{\tilde{\tau}}~,
\end{equation}
by the procedure given in the previous section. The overlap between any two coherent states is
\begin{equation}\label{AppCSOverlap}
\bra{\tau}\tau'\rangle=\frac{M(\tau,\tau')}{\sqrt{N(\tau)N(\tau')}}~,
\end{equation}
where
\begin{equation}
\begin{split}
M(\tau,\tau')&=\bra{l_0}e^{\tau^*\hat{K}_-}e^{\tau'\hat{K}_+}\ket{l_0}=\\
&=\sum_{n=0}^\infty\sum_{m=0}^\infty\frac{(\tau^*)^n}{n!}\frac{(\tau')^m}{m!}\bra{l_0}\hat{K}_-^n\hat{K}_+^m\ket{l_0}
\label{AppCS_M}
\end{split}
\end{equation}
is the overlap between two unnormalized GCS. Since $\ket{l_0}$ is the LWV, it holds that
\begin{equation}
\begin{split}
    \bra{l_0}[\hat{K}_-^n,\hat{K}_+^m]\ket{l_0}&=\bra{l_0}\hat{K}_-^n\hat{K}_+^m-\hat{K}_+^m\hat{K}_-^n\ket{l_0}=\\
    &=\bra{l_0}\hat{K}_-^n\hat{K}_+^m\ket{l_0}~,
\label{AppCommAdv}
\end{split}
\end{equation}
and hence
\begin{equation}
M(\tau,\tau')=\sum_{n=0}^\infty\sum_{m=0}^\infty\frac{(\tau^*)^n}{n!}\frac{(\tau')^m}{m!}\bra{l_0}[\hat{K}_-^n,\hat{K}_+^m]\ket{l_0}~.
\end{equation}
In any representation, the Lie bracket satisfies
\begin{equation}
    [AB,C]=A[B,C]+[A,C]B~,
\end{equation}
meaning that
\begin{equation}
\label{CSOverlap_1}
    \begin{split}
    \bra{l_0}&[\hat{K}_-^n,\hat{K}_+^m]\ket{l_0}=\\
    &=\bra{l_0}\hat{K}_+[\hat{K}_-^{n},\hat{K}_+^{m-1}]+[\hat{K}_-^{n},\hat{K}_+]\hat{K}_+^{m-1}\ket{l_0}=\\
&=\bra{l_0}[\hat{K}_-^{n},\hat{K}_+]\hat{K}_+^{m-1}\ket{l_0}~,
    \end{split}
\end{equation}
where we used Eq.~\eqref{e.HWV}. The commutator appearing above has the simple expression
\begin{equation}
[\hat{K}_+,\hat{K}_-^n]=\left(n\beta\hat{K}_0-\frac{n(n-1)}{2}\beta\alpha_-\right)\hat{K}_-^{n-1}~,
\label{e.fomula_by_induction}
\end{equation}
which can be shown by induction as follows. For $n=1$ the formula gives the representation of the first line of Eq.~\eqref{e.brackets}. Then, assuming the identity to hold true for $n-1$ gives 
Eq.~\eqref{e.fomula_by_induction}, as expected. In the same way one can show that
\begin{equation}
[\hat{K}_+^{n},\hat{K}_-]=\left(n\beta\hat{K}_0-\frac{n(n-1)}{2}\beta\alpha_+\right)\hat{K}_+^{n-1}~.
\end{equation}
Then, equation (\ref{CSOverlap_1}) gives
\begin{equation}
\bra{l_0}[\hat{K}_-^n,\hat{K}_+^m]\ket{l_0}=\Sigma(n)\bra{l_0}\hat{K}_-^{n-1}\hat{K}_+^{m-1}\ket{l_0}
\label{App_CSoverlap_to_be_proved}
\end{equation}
where
\begin{equation}
\Sigma(n)=\frac{n(n-1)}{2}\beta\alpha_--n\beta\nu_0~.
\end{equation}
By iterating this procedure, we find that (\ref{CSOverlap_1}) is nonzero only for $m=n$, for which it gives
\begin{equation}
\bra{l_0}\hat{K}_-^n\hat{K}_+^m\ket{l_0}=\delta_{n,m}\prod_{k=1}^n\Sigma(k)=\delta_{n,m}P_\mathfrak{g}^{\nu_0}(n)n!~, 
\end{equation}
where we introduced the function $P_\mathfrak{g}^{\pi,\nu_0}(n)$ which only depends on $n$, the algebra's structure constants, and the value of $\nu_0$ in the specific representation considered; its explicit expression is
\begin{equation}
    P_{\mathfrak{g}}^{\pi,\nu_0}(n)
    =(-1)^n\left(\frac{\beta\alpha_-}{2}\right)^n  (2\nu_0/\alpha_-)_n
\end{equation}
where $(x)_n$ is the Pochhammer symbol, defined by
\begin{equation}
    (x)_n=x(x-1)\dots(x-(n-1))~.
\end{equation}
In a similar fashion, by modifying Eq.~\eqref{e.Rep_LWV} so that $\ket{h_0}$ is the HWV annihilated by $\hat{K}_+$, one finds
\begin{equation}
    \bra{h_0}\hat{K}^n_+\hat{K}^m_-\ket{h_0}=\delta_{n,m}Q_{\mathfrak{g}}^{\pi,\nu_0}(n)n!
\end{equation}
with 
\begin{equation}
    Q_{\mathfrak{g}}^{\pi,\nu_0}(n)=(-1)^n\left(\frac{\beta\alpha_+}{2}\right)^n  (2\nu_0/\alpha_+)_n
\end{equation}

Finally, the the overlap of two Perelomov coherent states in the $\tau$ coordinates is
\begin{equation}
\bra{\tau}\tau'\rangle=\frac{1}{\sqrt{N(\tau)N(\tau')}}\sum_{n=0}^{\infty}\frac{(\tau^*\tau')^n}{n!}P_{\mathfrak{g}}^{\nu_0}(n)
\label{e.result1}
\end{equation}
when the GCS are built from the LWV, and
\begin{equation}
\bra{\tau}\tau'\rangle=\frac{1}{\sqrt{N(\tau)N(\tau')}}\sum_{n=0}^{\infty}\frac{(\tau\tau'^*)^n}{n!}Q_{\mathfrak{g}}^{\nu_0}(n)
\label{e.result2}
\end{equation}
if they are built from the HWV; once the transformation $\tau(\Omega)$ is given, analogue results for the $\Omega$ coordinates can be obtained from the above by substitution.

\section{Examples}
\label{sec:5}
In this section, we apply our formula to some Lie algebras,
re-deriving known results for $\mathfrak{su}(2)$ and $\mathfrak{su}(1,1)$, and providing new ones for the $\mathfrak{sl}(2,\mathbb{R})$ subalgebras of Virasoro.

\subsection{\texorpdfstring{$\mathfrak{su}(2)$}{su(2)}}
Let us consider the real Lie algebra $\mathfrak{g}=\mathfrak{su}(2)$ spanned by the three elements $\{T_1,T_2,T_3\}$ satisfying the Lie brackets
\begin{equation}
    [T_i,T_j]=i\epsilon_{ijk}T_k~,
    \label{e.su(2)_comm}
\end{equation}
where $\epsilon_{ijk}$ is the Ricci-Levi-Civita tensor. This is the unique compact real form of $\mathfrak{sl}(2,\mathbb{C})$, which is a simple Lie algebra of rank one. By considering $T_3$ and the linear combinations $T_{\pm}=T_1\pm iT_2$, the commutation relations take the form
\begin{equation}
\begin{dcases}
    [T_3,T_\pm]=\pm T_\pm\\
    [T_+,T_-]=2T_3
\end{dcases}
\end{equation}
and the Cartan subalgebra is spanned by $T_3$. Hence, the displacement operators in some representation $r=(\pi,\mathcal{H})$ are obtained as
\begin{equation}
    D(\Omega)=e^{\Omega \hat{T}_+-\Omega^* \hat{T}_-}~,
\end{equation}
where $\hat{T}_\pm=\pi(T_\pm)$. When expressed in terms of the original algebra elements $\{\hat{T}_1,\hat{T}_2,\hat{T}_3\}$, the above operator becomes
\begin{equation}
    D(\Omega)=e^{2i(\mathrm{Im}(\Omega)\hat{T}_1+\mathrm{Re}(\Omega)\hat{T}_2)}~,
\end{equation}
making it clear that the displacement is the complex exponential of a real linear combination of the algebra elements, as it should in the convention \eqref{e.exp_map}.

\subsubsection{GCS in the \texorpdfstring{$j$}{j} representations}
We now consider the unitary irrep. of $\mathfrak{su}(2)$ labelled by the (half-)integer $j$, whose GCS are well-known and widely used in the literature (see e.g. Ref.~\cite{Combescure21} for an extensive review on the subject). From the structure constants, we get $\alpha_\pm=\pm 1$ and $\beta=2$, and the action of the diagonal element on the LWV $\ket{j,-j}$ gives $\nu_0=-j$. Therefore, by plugging
\begin{equation}
P_{\mathfrak{su}(2)}^{j,-j}(n)=(-1)^n(-2j)_n=(2j-n+1)_n~.
\end{equation}
in (\ref{AppCS_M}), we get 
\begin{equation}
\begin{split}
    M(\tau,\tau')&=\sum_{n=0}^{2j}\frac{(\tau^*\tau')^n}{n!}(2j-n+1)_n\\
    &=\sum_{m=-j}^{j}\binom{2j}{m+j}(\tau^*\tau')^{m+j}~.
\end{split}
\label{App_CSoverlap_SU2}
\end{equation}
for the overlap of two unnormalized coherent states in the $\tau$ coordinates. Finally, inserting (\ref{App_CSoverlap_SU2}) in (\ref{AppCSOverlap}) gives
\begin{equation}
\bra{\tau}\tau'\rangle=\frac{(1+\tau^*\tau')^j}{(1+|\tau|^2)^j(1+|\tau'|^2)^j}
\end{equation}
for the $\mathfrak{su}(2)$ coherent states in the projective coordinates $\tau$, and
\begin{equation}
\bra{\Omega}\Omega'\rangle=\left(\cos\frac{\theta}{2}\cos\frac{\theta'}{2}+\sin\frac{\theta}{2}\sin\frac{\theta'}{2}e^{i(\phi'-\phi)}\right)^{2j}
\end{equation}
in the $\Omega=(\theta, \phi)$ coordinates. Additionally, the above result can be used to show that
\begin{equation}
    \lim_{j\to\infty}\lvert\bra{\Omega}\Omega'\rangle\rvert=\delta(\Omega-\Omega')~,
\end{equation}
meaning that the spin GCS becomes orthonormal in the semiclassical large-$j$ limit~\cite{ZyczkowskiS01}.

\subsection{\texorpdfstring{$\mathfrak{su}(1,1)$}{su(1,1)}}
Next, let us consider $\mathfrak{g}=\mathfrak{su}(1,1)$, which is the real Lie algebra spanned by the three elements $\{K_1,K_2,K_3\}$ satisfying the Lie brackets
\begin{equation}
    \begin{dcases}
        [K_0,K_1]=iK_2\\
        [K_0,K_2]=iK_1\\
        [K_1,K_2]=-iK_0
    \end{dcases}~.
    \label{e.su(1,1)_comm}
\end{equation}
This algebra is (isomorphic to) the split non-compact real form of $\mathfrak{sl}(2,\mathbb{C})$. By considering the linear combinations $K_{\pm}=K_1\pm iK_2$ and $K_0$, the commutation relations take the form
\begin{equation}
\begin{dcases}
    [K_0,K_\pm]=\pm K_\pm\\
    [K_+,K_-]=-2K_0
\end{dcases}~,
\label{e.su(1,1)_comm_pm}
\end{equation}
and the Casimir element is $C=K_0^2-\frac{1}{2}\{K_+,K_-\}$. The Cartan subalgebra is spanned by $K_0$, and hence the displacement operators in some representation $r=(\pi,\mathcal{H})$ are obtained as
\begin{equation}
    \hat{D}(\Omega)=e^{\Omega \hat{K}_+-\Omega^* \hat{K}_-}~,
\end{equation}
where $\hat{K}_\pm=\pi(K_\pm)$. The GCS of $\mathfrak{su}(1,1)$, sometimes called pseudo-spin CS, have many applications, from describing squeezing in quantum optics~\cite{RosasOrtiz19, KhannaEtAl09} to realising the thermofield double state in two copies of a CFT~\cite{CottrellEtAl19}, which is a main ingredient for many results in holography~\cite{Israel76, Maldacena03, Coppo23, ChapmanEtAl19, HartmanM2013}. $\mathfrak{su}(1,1)$ has many representations; in the following, we focus on those of greater interest for physics, namely the one- and two-mode representations.

\subsubsection{GCS in the two-mode representations}
The two-mode representation of $\mathfrak{su}(1,1)$ is defined by the operators
\begin{equation}
    \begin{dcases}
        \hat{K}_+=\hat{a}^\dagger \hat{b}^\dagger\\
        \hat{K}_-=\hat{a}\hat{b}\\
        \hat{K}_0=\frac{1}{2}(\hat{a}^\dagger\hat{a}+ \hat{b}^\dagger\hat{b}+1)
    \end{dcases}
\end{equation}
acting on two copies of a bosonic Fock space. The Casimir operator shows that the two-mode representations are labelled by  $k=(n_0+1)/2$ for $n_0\in\mathbb{N}$. Choosing the LWV
\begin{equation}
    \ket{\psi_0}=\ket{n_0}\otimes\ket{0}
\end{equation}
as the reference for the GCS construction gives $\alpha_\pm=\pm 1$, $\beta=-2$ and $\nu_0=k$, and hence
\begin{equation}
P_{\mathfrak{su}(1,1)}^{k}(n)=(n_0+1)_n=\frac{(n_0+n)!}{n_0!}~.
\end{equation}
The function $M(\tau,\tau')$ is
\begin{equation}
M(\tau,\tau')=\sum_{n=0}^\infty \frac{(n_0+n)!}{n_0!n!}(\tau^*\tau')^n
\end{equation}
and the overlap results in
\begin{equation}
\bra{\tau}\tau'\rangle=(1-|\tau|^2)^{k}(1-|\tau'|^2)^{k}\sum_{n=0}^{\infty}\frac{(n_0+n)!}{n_0!n!}(\tau^*\tau')^{n}~.
\end{equation}
Since the sum
\begin{equation}
\sum_{n=0}^\infty \frac{(n_0+n)!}{n_0!n!}x^n= \frac{1}{(1-x)^{(1+n_0)}}
\end{equation}
converges when $x=\tau^*\tau'$ is a point in the Poincar\'e disk, we obtain
\begin{equation}
\bra{\tau}\tau'\rangle=\frac{(1-|\tau|^2)^{k}(1-|\tau'|^2)^{k}}{(1-\tau^*\tau')^{2k}}
\label{e.su(1,1)_result}
\end{equation}
in $\tau$ coordinates. Again, the result we obtained can be used to show that
\begin{equation}
    \lim_{k\to\infty}\lvert\bra{\tau}\tau'\rangle\rvert=\delta(\tau-\tau')~,
    \label{e.semi-classi_su(1,1)}
\end{equation}
meaning that the pseudo-spin CS becomes orthonormal in the semiclassical large-$k$ limit~\cite{CoppoEtAl20}.

\subsubsection{GCS in the one-mode representations}
The one-mode representation of $\mathfrak{su}(1,1)$ is defined by the operators
\begin{equation}
    \begin{dcases}
        \hat{K}_+=\frac{1}{2}(\hat{a}^\dagger)^2\\
        \hat{K}_-=\frac{1}{2}\hat{a}^2\\
        \hat{K}_0=\frac{1}{4}(\hat{a}\hat{a}^\dagger+\hat{a}^\dagger\hat{a})
    \end{dcases}
\end{equation}
acting on a bosonic Fock space. The Casimir operator shows that there are two one-mode representations, respectively labelled by  $k=1/4$ and $k=3/4$. These correspond to the lowest-weight vectors
\begin{equation}
    \ket{\psi^{(1/4)}_0}=\ket{0}~,~~\ket{\psi^{(3/4)}_0}=\ket{1}
\end{equation}
which can be used as the extremal state vector for the GCS construction. The two representations are called the even and odd one-mode representations~\cite{Novaes04}. The two lowest weight vectors $\ket{\psi^{(k)}_0}$ satisfy
\begin{equation}
    \hat{K}_0\ket{\psi^{(k)}_0}=k\ket{\psi^{(k)}_0}~;
\end{equation}
hence, $\alpha_\pm=\pm 1$, $\beta=-2$ and $\nu_0=k$. Therefore, 
\begin{equation}
P_{\mathfrak{su}(1,1)}^{k}(n)=(2k)_n=\begin{dcases}
    (1/2)_n~&{\rm if~}k=1/4\\
    (3/2)_n~&{\rm if~}k=3/4
\end{dcases}
\end{equation}
In the first case,
\begin{equation}
    (1/2)_n=\frac{\Gamma(n+1/2)}{\Gamma(1/2)}=\frac{2(2n-1)!}{2^{2n}(n-1)!}
\end{equation}
gives
\begin{equation}
    M(\tau,\tau')=
    \frac{1}{\sqrt{1-\tau^*\tau'}}~;
\end{equation}
similarly, in the second case, 
\begin{equation}
    (3/2)_n=\left(2n+1\right)(1/2)_n=\frac{(2n+1)!}{2^{2n}n!}
\end{equation}
gives
\begin{equation}
    M(\tau,\tau')=\frac{1}{(1-\tau^*\tau')^{3/2}}~;
\end{equation}
In both cases, we obtain the same expression as in Eq.~\eqref{e.su(1,1)_result}, as well as the same semi-classical limit.

\subsection{The \texorpdfstring{$\mathfrak{sl}(2,\mathbb{R})$}{sl(2,R)} subalgebras of Virasoro}
Finally, let us consider the Virasoro algebra defined by the commutation relations
\begin{equation}
    [L_m,L_n]=(m-n)L_{m+n}+\frac{c}{12}m(m^2-1)\delta_{m,-n}~.
\end{equation}
The relevance of this algebra is that it describes the symmetries of a two-dimensional conformal field theory, where $c$ plays the role of the CFT's central charge~\cite{DiFrancescoEtAl97}. As is clear from the above Lie brackets, the subsets $\mathfrak{g}_k=\{L_{-k},L_0,L_k\}$ are closed subalgebras. Specifically, these have commutation relations
\begin{equation}
\begin{dcases}
    &[L_0,L_{\pm k}]=\mp k L_{\pm k}\\
    &[L_{+k},L_{-k}]=2kL_0+\frac{c}{12}k(k^2-1)
\end{dcases}~.
\end{equation}
In Ref.~\cite{CaputaG23}, the algebras $\mathfrak{g}_k$ were expressed as copies of $\mathfrak{sl}(2,\mathbb{R})$ and analysed for their relevance in holography and the AdS/CFT correspondence. As $\mathfrak{sl}(2,\mathbb{R})$ is isomorphic to $\mathfrak{su}(1,1)$~\cite{Gilmore12}, here we define
\begin{equation}
\begin{dcases}
    \mathcal{L}_0=-\frac{1}{k}\left(L_0+\frac{c}{24}(k^2-1)\right)\\
    \mathcal{L}_{\pm}=-\frac{1}{k}L_{\pm k}
\end{dcases}
\end{equation}
and express $\mathfrak{g}_k$ as $\mathfrak{su}(1,1)$ algebras identified by $k$ and having commutation relations
\begin{equation}
\begin{dcases}
    [\mathcal{L}_0,\mathcal{L}_\pm]=\pm\mathcal{L}_\pm\\
    [\mathcal{L}_+,\mathcal{L}_-]=-2\mathcal{L}_0
\end{dcases}~.
\end{equation}
In CFT, one builds highest-weight vectors $\ket{h}$ by the action of an (anti-)holomorphic field of weight $h$ on the $SL(2,\mathbb{R})$ invariant vacuum $\ket{0}$. By construction, these HWVs satisfy $\hat{L}_0\ket{h}=h\ket{h}$ and $\hat{L}_{k}\ket{h}=0,~\forall k>0$. Therefore, we can build the coherent states
\begin{equation}
    \ket{\Psi_k(\xi)}=e^{\xi\hat{L}_{-k}-\xi^*\hat{L}_{+k}}\ket{h}~,
    \label{e.vir_GCS}
\end{equation}
and apply our formula to obtain the overlap of these GCS. First, we rewrite Eq.~\eqref{e.vir_GCS} as
\begin{equation}
    \ket{\xi;k}=e^{\xi^* k\hat{\mathcal{L}}_k-\xi k\hat{\mathcal{L}}_{-k}}\ket{h}~,
\end{equation}
and notice that
\begin{equation}
    \hat{\mathcal{L}}_0\ket{h}=-\frac{1}{k}\left(h+\frac{c}{24}(k^2-1)\right)\ket{h}\equiv -h'\ket{h}~.
\end{equation}
Therefore, Eq.~\eqref{e.vir_GCS} defines the GCS of $\mathfrak{su}(1,1)$ parameterised by $\Omega=k\xi^*$, and built from the HWV with weight $-h'$, where $h'\in\mathbb{R}^+$. For the sake of simplicity, we introduce $\Omega=\rho e^{i\phi}$ and $\xi=R e^{i\delta}$, meaning that $\rho=kR$ and $\phi=-\delta$. In App.~\ref{app:Tau-Omega}, we show how to obtain the $\tau$-coordinates expression
\begin{equation}
    \ket{\tau}=N^{-1/2}(\tau)e^{-\tau^*\hat{\mathcal{L}}_{-k}}\ket{h}~,
\end{equation}
for the GCS of $\mathfrak{su}(1,1)$ built from an HWV, where
\begin{equation}
    \tau=e^{i\phi}\tanh{\rho}=e^{-i\delta}\tanh(kR)
\end{equation}
and
\begin{equation}
    N^{1/2}(\tau)=\frac{1}{\cosh^{-2h'}(\rho)}=\cosh^{2h'}(k R)~.
\end{equation}
Thanks to our formula, the fact that
\begin{equation}
    Q^{h'}_{\mathfrak{su}(1,1)}(n)=(2h')_n~,
\end{equation}
and the previous results about $\mathfrak{su}(1,1)$, it is easy to show that
\begin{equation}
    \begin{split}
        \bra{\xi;k}\xi';k\rangle&=\frac{\sum_{n=0}^\infty \frac{(\tau(\xi)\tau(\xi')^*)^n}{n!} Q^{h'}_{\mathfrak{su}(1,1)}(n)}{\sqrt{N(\tau(\xi))N(\tau(\xi'))}}=\\
        &=\left[\frac{1-\tanh{(kR)}\tanh{(kR')}e^{i(\delta'-\delta)}}{\cosh(kR)\cosh(kR')}\right]^{2h'}~.
    \end{split}
\end{equation}

As it is clear, at all fixed $k>1$ there are two independent parameters driving the semi-classical limit: namely, the central charge $c$ of the CFT and the weight $h$ of the (anti-)holomorphic field used to generate the HWV. Using Eq.~\eqref{e.semi-classi_su(1,1)}, it is easy to show that
\begin{equation}
    \lim_{c\to\infty}\lvert\bra{\xi;k}\xi';k\rangle\rvert=\delta(\xi-\xi')
\end{equation}
and
\begin{equation}
    \lim_{h\to\infty}\lvert\bra{\xi;k}\xi';k\rangle\rvert=\delta(\xi-\xi')~,
\end{equation}
meaning that we can approach the semi-classical limit by either increasing $c$ or picking larger a $h$ for the GCS reference; while the second is equivalent to what we encountered in all the above examples, the first is a specific feature of the Virasoro algebra coming from the fact that it allows tuning its structure constants by changing $c$. Notice that the dependence of $h'$ on $c$ disappears at $k=1$, meaning that the Virasoro subalgebra $\mathfrak{g}_1$ becomes semiclassical at large $h$ only. Moreover, assuming there is an observer-dependent threshold $h'_{\rm t}$ that must be reached so that the CFT looks effectively classical for that observer~\cite{CoppoEtAl22}, then classicality can be achieved by either tuning $c$ and $h$: a CFTs in some $\mathfrak{g}_{k>1}$-GCS looks effectively classical whenever it is identified by a point $(h,c)$ above the line
\begin{equation}
    h(c)=kh'_{\rm t}-\frac{c}{24}(k^2-1)
\end{equation}
in the $h-c$ parameter space.

\section{Conclusions}
\label{sec:6}
In this paper, we provided a formula for computing the overlap of two Generalized Coherent States (GCS) of any rank one Lie algebra admitting triangular decomposition, e.g. any simple rank one Lie algebra. The result, contained in Eq.s (\ref{e.result1}) and (\ref{e.result2}), is a compact expression that only requires summing over powers of the complex numbers identifying the CGS multiplied with simple functions that depend on the algebra's structure constants and on the action of the Cartan subalgebra on the extremal vector used for the GCS construction. 

Proceeding further, generalising our formula to semi-simple Lie algebras of any rank would be the next logical step. However, this comes with the additional technical complication associated with controlling the nested commutators appearing in the BCH formula. As a starting point, the generalisation to rank two semi-simple Lie algebras looks more manageable, as the information encoded in the root space decomposition may be used to obtain a two-dimensional analogue of the formula presented herein.

In conclusion, the formula presented in this article may find broad applicability in many settings, from studying the quantum-to-classical crossover to analysing the properties of semi-classical quantum systems, from investigating the complexity and large-$N$ limit of coherent CFTs to uncovering the relation between the classical and quantum behaviours of black holes. Moreover, our formula can compute the fidelity and other information-theoretic quantities relative to coherent and squeezed states accessible in the lab, and hence be used for the theoretical analysis of a wide class of experiments. Finally, our formula is the first step of a realization-independent characterization of which are the essential classical-like features of a GCS construction, hence paving the way for studying the emergence of the classical world from the quantum one in a model-independent manner.

\section*{Acknowledgments}
The author acknowledges financial support from the Magnus Ehrnrooth Foundation and the Academy of Finland via the Centre of Excellence program (Project No. 336810 and Project No. 336814).

\bibliography{bib}

\appendix

\section{The unnormalized \texorpdfstring{$\mathfrak{su}(1,1)$}{su(1,1)} GCS built from a HWV}
\label{app:Tau-Omega}
In this appendix, we show how to obtain the $\tau$ coordinates for the GCS of $\mathfrak{su}(1,1)$ built from a HVW $\ket{h}$ by the BCH formula. To this end, let us consider the so-called split basis representation, in which
\begin{align}
        K_0&=\frac{1}{2}\begin{pmatrix}
            1 & 0 \\
            0 & -1
        \end{pmatrix}~,\\
        K_+&=\begin{pmatrix}
            0 & i \\
            0 & 0
        \end{pmatrix}~,\\
        K_-&=\begin{pmatrix}
            0 & 0 \\
            i & 0
        \end{pmatrix}~.
    \end{align}
Hence
\begin{equation}
    K_+^2=K_-^2=0~,
    \label{a.su_squares}
\end{equation}
and
\begin{equation}
    \{K_+,K_-\}=-\mathbb{I}~.
    \label{a.su_anticomm}
\end{equation}
The GCS are obtained by the action of
\begin{equation}
    D(\Omega)=e^{\Omega K_+-\Omega^* K_-}
    \label{a.su_displacement}
\end{equation}
over the HWV annihilated by $K_+$, where $\Omega\in\mathbb{C}$. To get the expression of the unnormalized GCS, we note that Eq.s~\eqref{a.su_squares}-\eqref{a.su_anticomm} imply
\begin{equation}
    (\Omega K_+-\Omega^* K_-)^{2n}=|\Omega|^{2n}\mathbb{I}
\end{equation}
 and
\begin{equation}
    (\Omega K_+-\Omega^* K_-)^{2n+1}=|\Omega|^{2n}(\Omega K_+-\Omega^*K_-)~;
\end{equation}
consequently, the definition of the exponential as a series expansion gives
\begin{equation}
    e^{\Omega K_+-\Omega K_-}=\mathbb{I}\cosh|\Omega|+\frac{\sinh|\Omega|}{|\Omega|}\left(\Omega K_+-\Omega^*\Omega K_-\right)~.
\end{equation}
Hence, we obtain the matrix expression of the displacement operator as
    \begin{equation}
        D(\Omega)=\begin{pmatrix}
            \cosh|\Omega| & i\frac{\sinh|\Omega|}{|\Omega|}\Omega\\
            -i\frac{\sinh|\Omega|}{|\Omega|}\Omega^* & \cosh|\Omega|
            \end{pmatrix}
            \label{a.matrix1}
    \end{equation}
or, equivalently,
    \begin{equation}
        D(\Omega)=\begin{pmatrix}
            \cosh\rho & ie^{i\phi}\sinh\rho\\
            -ie^{-i\phi}\sinh\rho & \cosh\rho
            \end{pmatrix}
            \label{a.su_matrix1}
    \end{equation}
where $\Omega=\rho e^{i\phi}$. Next, we use the BCH formula to write 
\begin{equation}
        D(\Omega)=e^{-\tau^* X_-}e^{-\eta H}e^{\tau X_+}~,
\end{equation}
for some $\tau$ and $\eta$. Since
\begin{equation}
    e^{\tau X_+}=\mathbb{I}+\tau X_+~,~~e^{-\tau^* X_-}=\mathbb{I}-\tau^* X_-~,
\end{equation}
and
\begin{equation}
    e^{-\eta H}=\begin{pmatrix}
        e^{-\eta/2} & 0\\
        0 & e^{\eta/2}
    \end{pmatrix}~,
\end{equation}
we get
    \begin{equation}
     D(\Omega)=\begin{pmatrix}
         e^{-\eta/2}&  i\tau e^{-\eta/2}\\
         -i\tau^*e^{-\eta/2}  & e^{\eta/2}+||\tau||^2e^{-\eta/2} 
    \end{pmatrix}~.
    \label{a.su_matrix2}
\end{equation}
Next, we compare \eqref{a.su_matrix1} and \eqref{a.su_matrix2}, and get
\begin{equation}
    \tau(\Omega)=e^{i\phi}\tanh\rho
\end{equation}
and
\begin{equation}
    \eta(\Omega)=\ln\left(\frac{1}{\cosh^2\rho}\right)~.
\end{equation}

To get the unnormalized GCS in some given representation, let us consider the HWV defined by
\begin{equation}
    K_+\ket{\psi_0}=0
\end{equation}
and
\begin{equation}
    K_0\ket{\psi_0}=h\ket{\psi_0}~.
\end{equation}
Then,
\begin{equation}
    \begin{split}
        \ket{\Omega}&=\hat{D}(\Omega)\ket{\psi_0}\\
        &=e^{-\tau^* K_-}e^{-\eta K_0}\ket{\psi_0}=N^{-1/2}(\tau)\ket{\Tilde{\tau}}~;
    \end{split}
\end{equation}
therefore,
\begin{equation}
    \ket{\Tilde{\tau}}=e^{-\tau^* X_-}\ket{\psi_0}
\end{equation}
and
\begin{equation}
    N^{-1/2}(\tau)=\cosh^{2h}(\rho)~.
\end{equation}
\end{document}